# Highly efficient CdTe solar cell with a thin CIT current booster: theoretical insights

Md. Sabbir Hossain[1], Md. Alamin Hossain pappu[1], Bipanko Kumar Mondal[2], Ahnaf Tahmid Abir[1], and Jaker Hossain[1*]

*[1]Solar Energy Laboratory, Department of Electrical and Electronic Engineering, University of Rajshahi, Rajshahi 6205, Bangladesh.*

*[2]Department of Electrical & Electronic Engineering, Pundra University of Science & Technology, Bogura, Bogura 5800, Bangladesh.*

**Abstract**

CdTe-based thin film solar cell has been modeled and enumerated with a thin $CuInTe_2$ (CIT) current booster layer. CdTe-based *n*-CdS/*p*-CdTe/$p^+$-CIT/$p^{++}$-$WSe_2$ heterojunction device has been evaluated for the highest performance. It is revealed that physical parameters such as thickness, doping, and defects of the CIT layer have a significant influence on the performance of the CdTe solar cell. The device shows an efficiency of 37.46% with an open circuit voltage, $V_{OC}$ of 1.102 V, short circuit current density, $J_{SC}$ of 38.50 mA/cm$^{2,}$ and fill factor, FF of 88.30%. The use of the photon recycling technique with a Bragg-reflector with 98% back and 95% front reflectance only provides an efficiency of ~44.3% with a current of 45.4 mA/cm$^2$. These findings are very hopeful for the production of an efficient CdTe solar cells in the near future.

## 1. Introduction

Cadmium telluride (CdTe)-based thin film photovoltaic (PV) cells are one of the most prominent and fruitful PV technologies owing to their cost effectiveness, high efficiency, reliability and stability [1]. CdTe, as a part of II-VI family, is nearly an ideal semiconductor to manufacturing thin film solar cells as possesses a bandgap of 1.5 eV and higher optical absorption coefficient, α of ~$10^5$ $cm^{-1}$ [2]. Cadmium sulfide (CdS), another II-VI material, has been utilized as an n-type partner with p-type CdTe to form a suitable p-n junction because of its wide bandgap of 2.4 eV and diffusion of S to form a ternary $CdTe_{1-x}S_x$ at the interface which reduces structural defects and hence recombination current in the device [1]. In 2016, First Solar Company announces high performance CdTe solar cells with a record highest power conversion efficiency (PCE) of 22.1% maintaining an open circuit voltage ($V_{OC}$) of 0.887 V and short circuit voltage ($J_{SC}$) of 31.69 mA/$cm^2$ [3].

In photovoltaic cells, the open circuit voltage ($V_{OC}$) rises and the short circuit voltage ($J_{SC}$) falls when the absorber bandgap elevates [4-5]. Therefore, the optimal bandgap for the highest efficiency presents the balance between $J_{SC}$ and $V_{OC,}$ and the maximum efficiency can be achieved for the absorber bandgap of around 1.4-1.5 eV [6-7]. In that sense, CdTe could provide very high efficiency compared to Si solar cells. It offers a high $V_{OC}$, however, the $J_{SC}$ of the solar cell is lower just because of the high bandgap of the material which impedes generating high efficiency of the device. On the contrary, CdTe solar cells fabricated by First Solar Company attained a $J_{SC}$ as mentioned earlier that certainly exceeds the Shockley-Queisser (SQ) limit for $J_{SC}$ of 26 mA/$cm^2$ [8]. Therefore, there is a still hope to increase the $J_{SC}$ of the device which will ameliorate the efficiency of the CdTe solar PV devices. However, some mechanisms are already proposed for the rise in short circuit current in CdTe solar cells such as the use of a CdSe window layer for the absorption of shorter and longer wavelength lights [1, 9], impurity PV (IPV) effect in which electron-hole pairs are produced by adding low energy photons involving impurity in the CdTe absorber layer, impact ionization effect in graded bandgap solar cells, the introduction of electron back diffusion barrier (ebdb) and hole back diffusion barrier (hbdb) layers in the cells, etc. [8, 10].

However, in this work, we propose a different technique with a thin and relatively narrow bandgap $CuInTe_2$ layer that along with the $WSe_2$ back surface field (BSF) layer that works as a current booster layer in CdTe solar cells.

$CuInTe_2$ is an I-III-VI group semiconducting material obtaining a direct bandgap of 0.9-1.1 eV [11-12]. It possesses a higher absorption coefficient of $10^5$ $cm^{-1}$ and defect tolerance that make it a superior candidate to be used in photodetectors [13], photovoltaic cells [11-12], and light-emitting diodes [14] etc. Moreover, $CuInTe_2$ thin film can readily be fabricated by various methods such as thermal deposition, molecular beam epitaxy (MBE), solvothermal, electrodeposition, etc. [11]. Therefore, CIT layer can be used as a potential absorber layer for low energy photons in CdTe solar cells.

In addition, CdS is an appropriate semiconductor to be exerted as a window material in different types thin film photovoltaic cells due to its high transmittance and stability [11]. On the other hand, $WSe_2$ is a chalcogenide semiconductor with an optical bandgap of 1.6-1.65 eV [15]. It is an influential candidate for photoelectrochemical and solar cells. Several methods are available for the manufacture of highly improved $WSe_2$ thin films such as CVD, drop cast, salinization method, and low-temperature CVD [15-16]. Therefore, CdS and $WSe_2$ are suitable candidates as the window and BSF layers for the preparation of high quality CdTe heterojunction thin film solar cells.

In this attempt, CdTe-based high efficiency *n*-CdS/*p*-CdTe/$p^+$-CIT/$p^{++}$-$WSe_2$ thin film solar device has been reported where a thin CIT layer acts as a current booster along with $WSe_2$ semiconducting layer which functions as a BSF layer material. The modeled PV device has been computed with solar cell capacitance simulation (SCAPS-1D) software for the highest performance. The computation findings indicate the great ability for the construction of high-power generating CdTe-based heterojunction thin film solar cells in the future.

## 2. Cell architecture and computer simulation

The CdTe solar device architecture and lighted electronic energy structure are shown in Figure 1(a) and (b), respectively. It consists of a window of n-type CdS, p-type CdTe absorber, $p^+$-type CIT current booster, and $p^{++}$-type $WSe_2$ BSF layers. The symbols $p^+$ and $p^{++}$ are used to indicate subsequent higher doping in this work. The CdTe with an optical bandgap of 1.5 eV owns an electron affinity of 4.28 eV and, on the other side, CdS possesses an optical bandgap and electron affinity of 2.4 and 4.2 eV, respectively. Therefore, CdS can build a type II heterojunction with CdTe. On the other hand, CIT semiconductor has a bandgap of 1.1 eV and electron affinity of 4.28 eV which enables it to form a suitable heterojunction with CdTe. Now, $WSe_2$ semiconductor seizes an optical bangap of 1.65 eV and electron affinity of 3.7 eV and as a consequence it can form a suitable BSF heterojunction with CIT layer as well as CdTe layer. Ni and Al metals have been taken as anode and cathode contacts for the collection of holes and electrons efficiently from the device.

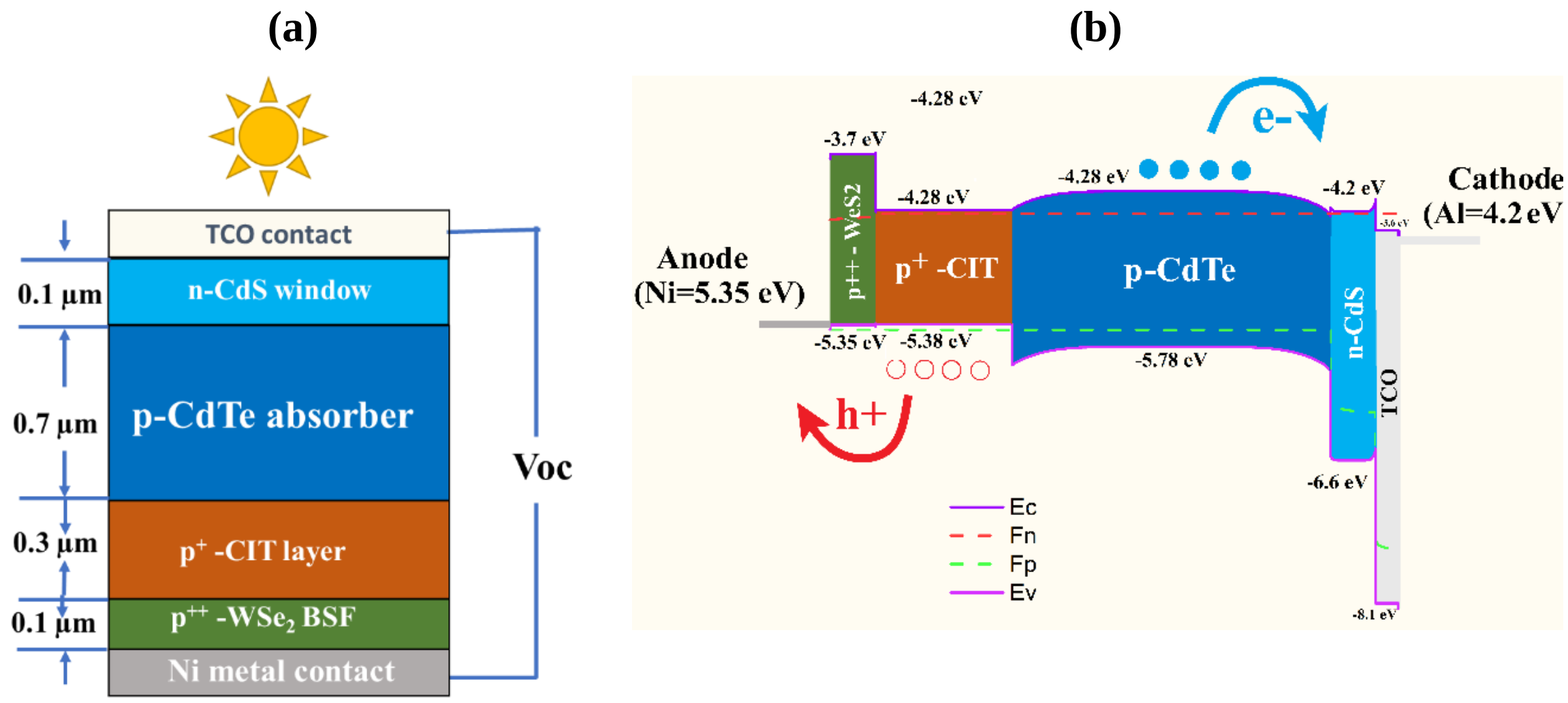

**Figure 1:** (a) The sketch and (b) alighted electronic band alignment of *n*-CdS/*p*-CdTe/$p^+$-CIT/$p^{++}$-$WSe_2$ photovoltaic cell.

**Table 1:** Different parameters conducted for the calculation of *n*-CdS/*p*-CdTe/$p^+$-CIT/$p^{++}$-$WSe_2$ thin film photovoltaic cells at the operating temperature of 300 K.

| **Parameters** | **ITO [17]** | **n-CdS [11]** | **p-CdTe [16]** | **$P^+$-CIT [11]** | **$P^{++}$-$WSe_2$ [15]** |
|---|---|---|---|---|---|
| Thickness (µm) | 0.05 | 0.1* | 0.7* | 0.3* | 0.1* |
| Band gap (eV) | 3.60 | 2.40 | 1.50 | 1.10 | 1.65 |
| Electron affinity (eV) | 4.50 | 4.20 | 4.28 | 4.28 | 3.70 |
| Dielectric constant | 8.90 | 9.35 | 10.30 | 11.0 | 13.8 |
| Effective DOS at CB (1/cm$^3$) | $2.2\times10^{18}$ | $2.2\times10^{18}$ | $9.20\times10^{17}$ | $5.08\times10^{16}$ | $8.30\times10^{18}$ |
| Effective DOS at VB (1/cm$^3$) | $1.8\times10^{19}$ | $1.8\times10^{19}$ | $5.20\times10^{18}$ | $5.65\times10^{18}$ | $1.60\times10^{19}$ |
| Mobility of electrons (cm²/Vs) | $5.0\times10^{1}$ | $1.0\times10^{2}$ | $3.2\times10^{2}$ | $3.0\times10^{3}$ | $1.0\times10^{2}$ |
| Mobility of holes (cm²/Vs) | $1.0\times10^{1}$ | $2.5\times10^{1}$ | $4.0\times10^{1}$ | $2.0\times10^{1}$ | $5.0\times10^{2}$ |
| Donor density, $N_D$ (1/cm$^3$) | $1.0\times10^{21}$ | $1.0\times10^{18}$ | - | - | - |
| Acceptor density, $N_A$ (1/cm$^3$) | $1.0\times10^{7}$ | - | $1.0\times10^{16}$ | $5.0\times10^{19}$* | $1.0\times10^{20}$* |
| Radiative recombination coefficient (cm$^3$/s) | $2.3 \times 10^{-11}$ | $2.3 \times 10^{-9}$ | $2.3 \times 10^{-9}$ | - | $1.04\times10^{-10}$ |
| Defect density (cm$^{-3}$) (above $E_v$ w.r.t. $E_{ref}$ (eV)) | - | 1.2 | 0.75 | 0.55 | 0.85 |
| | $1.00\times10^{10}$ | $1.00\times10^{14}$ | $1.00\times10^{14}$ | $1.00\times10^{14}$ | $1.00\times10^{15}$ |

**Interface defect parameters**

| Parameters | CdS/CdTe interface | CdTe/CIT interface | CIT/$WSe_2$ interface |
|---|---|---|---|
| Defects (cm$^{-2}$) | $1.00\times10^{11}$ | $1.00\times10^{11}$ | $1.00\times10^{11}$ |
| Defect type | Neutral | Neutral | Neutral |

*variable

The devices were analyzed with SCAPS-1D simulator that calculate $J_{SC}$, $V_{OC}$, FF and PCE of the solar PV cell solving Poison’s, continuity and drift-diffusion equations of carriers [18-19].

The single sun (100 mW/cm$^2$), AM 1.5G standard spectrum, optimal series and shunt resistances, and 300 K as the working temperature were considered during the entire computation. The thermal velocity of electron/hole of $10^7$ cm/s have been chosen for each layer to perform this calculation. For both metallic contacts, the recombination velocity at the surface was accounted at $10^7$ cm/s. The radiative recombination coefficients were used for each layer as noted on Table 1. Acceptor and donor defects were set to have a Gaussian energetic distribution for core layers. The SCAPS's absorption model served as the source of the absorption coefficient for every layer. A number of published data were used to provide the physical characteristics of each layer as specified in Table 1.

## 3. Results and discussion

### 3.1 CdTe PV cell with CIT current booster

In this section, the photovoltaic properties of *n*-CdS/*p*-CdTe/$p^+$-CIT/$p^{++}$-$WSe_2$ thin film PV cell have been illustrated in details. In the structure, CdS, CdTe, CIT and $WSe_2$ layers have thicknesses of 0.1, 0.7, 0.3 and 0.1 µm, in turn. The doping densities of the corresponding semiconductors are $10^{18}$, $10^{16}$, $5\times10^{19}$ and $10^{20}$ $cm^{-3}$, respectively. The bulk defects in each layer have important impacts on the performance of the solar cells. Therefore, reasonable amounts of bulk defects of $10^{14}$, $10^{14}$, $10^{14}$ and $10^{15}$ $cm^{-3}$ have been considered for CdS, CdTe, CIT and $WSe_2$ layers, respectively.

Figure 2(a) shows a direct comparison of the open circuit voltage, $V_{OC}$ and short circuit current, $J_{SC}$ of four combinations of CdTe photovoltaic cells. The *n*-CdS/*p*-CdTe configuration has the $V_{OC}$ of 0.987 volts, $J_{SC}$ of 25.2 mA/$cm^2$, FF of 82.14% and an efficiency of 20.43%.

The short circuit current density, $J_{SC}$ relies on the bandgap, $E_g$ of the absorber layer, and can be computed from the following equation [20]

$$J_{SC} = q\int_{E_g}^{\infty} \phi(\lambda)\mathrm{IQE}(\lambda)d\lambda \tag{1}$$

where, q denotes the electronic charge, $\phi(\lambda)$ stands for spectrum of the incoming light (here, AM 1.5G spectrum) and $\mathrm{IQE}(\lambda)$ is the internal quantum efficiency.

The minimum saturation current $J_0$ of a diode can be given by [21]

$$J_0 = \frac{q}{K_B}\frac{15\sigma}{\pi^4}T^3\int_u^\infty \frac{x^3}{e^x-1}dx \tag{2}$$

where, σ represents the Stefan–Boltzmann constant, and $u = \frac{E_g}{K_B T}$.

The open circuit voltage, $V_{OC}$ of a device can be found from the following formula [20]

$$V_{OC} = \frac{nK_BT}{q}ln\left(\frac{J_{SC}}{J_0}+1\right) \tag{3}$$

where, $K_B$ denotes the Boltzmann constant, T denotes the temperature.

The absorber bandgap dependent $V_{OC}$, $J_{SC}$ and dark current density of a device are shown in Figure 2(b) for AM 1.5G spectrum. The lower value of $J_{SC}$ is produced in the CdTe solar device because of the wide bandgap of CdTe semiconductor layer. However, it provides a $V_{OC}$ higher than Si or other relatively narrow bandgap solar cells due to higher bandap as can be seen from the figure. Therefore, the efficiency of the CdS/CdTe solar cell is limited to ~20.4%.

However, with the use of a thin $WSe_2$ back surface layer, the *n*-CdS/*p*-CdTe/$p^+$-$WSe_2$ PV device configuration shows a significant improvement in the voltage, $V_{OC}$ with a little increment in current, $J_{SC}$. The PV cell has the $V_{OC}$ of 1.125 Volt, FF of 88.73% and $J_{SC}$ of 27.4 mA/cm$^2$ and a PCE of 27.35%

The development of open circuit voltage, $V_{OC}^{pp^{++}}$ due to $pp^{++}$ junction can be given by [22]

$$V_{OC}^{pp^{+}} \cong \frac{K_BT}{q}ln\left[\frac{D_n}{N_p(w_n+w_p+w_{p^{++}})}\int_0^{w_{p^{++}}}\frac{N_{P^{++}}}{D_n}dx\right] \tag{4}$$

where, $w_n$, $w_p$ and $w_{p^{++}}$ are the base, emitter and BSF width, respectively. $D_n$ is the electron diffusion coefficient.

The $V_{OC}^{pp^{++}}$ due to $pp^{++}$ junction can also be given by [23]

$$V_{OC}^{pp^{++}} = \frac{K_B T}{q} ln\left[\frac{(1+b)\delta P(0)}{P_0(0)} + 1\right] \tag{5}$$

Where, $b = \frac{\mu_n}{\mu_p}$ is the electron and hole mobilities ratio. $P_0(0)$ is the thermal equilibrium density of holes, $\delta P(0)$ is the variation in hole density in the depletion region of the $pp^{++}$ heterojunction.

The electron current density at the np heterojunction by the effect of $pp^{++}$ BSF is given by [24]

$$J_{n,photo} = \frac{qF_0(1+\alpha L_n)e^{-\alpha w_n}}{\alpha L_n(1-\frac{1}{\alpha^2 L_n^2})} - \frac{2qF_0 e^{-\frac{w_n}{L_n}}}{\alpha L_n\left(1-\frac{1}{\alpha^2 L_n^2}\right)} \times \left[\frac{\left(\frac{D_n}{S_{pp^{++}} L_n}+1\right)e^{\left[-\alpha+(\frac{1}{L_n})\right]w_n} + \alpha D_n\left(\frac{1}{S_{pp^{++}}}-\frac{1}{\alpha D_n}\right)e^{\left[-\alpha+(\frac{1}{L_n})\right]d}}{\left(\frac{D_n}{S_{pp^{++}} L_n}-1\right)e^{-2d/L_n} + \left(\frac{D_n}{S_{pp^{++}} L_n}+1\right)e^{-2w_n/L_n}}\right] \tag{6}$$

for $\alpha L_n \neq 1$,

where, $F_0$ denotes the incident photon flux at the cell, $d = w_n + w_p$ is the sum of base and emitter width, $L_n$ presents the electron diffusion length, α denotes the absorption coefficient, $S_{pp^{++}}$ is the leakage velocity of the minority carriers moving from $pp^{++}$ junction. $S_{pp^{++}}$ can be determined from the following equation,

$$S_{pp^{++}} = \frac{(D_{np^{++}})N_P}{(L_{np^{++}})N_{P^{++}}}\left[coth\left(\frac{w_{p^{++}}}{L_{np^{++}}}\right)\right] \tag{7}$$

where, $D_{np^{++}}$ and $L_{np^{++}}$ represent the diffusion coefficient and diffusion length, respectively of the minority electrons within the $p^{++}$ layer. $N_P$ and $N_{P^{++}}$ denote the doping densities inside the $p$ and $p^{++}$ layers, respectively.

Therefore, with the inclusion of the $p^{++}$-$WSe_2$ back surface semiconductor, the $V_{OC}$ of the CdTe solar cell enhances by 0.145 V and $J_{SC}$ of the device increases by 2.2 mA/cm$^2$ owing to the operation of the built-in electric field developed at the back surface.

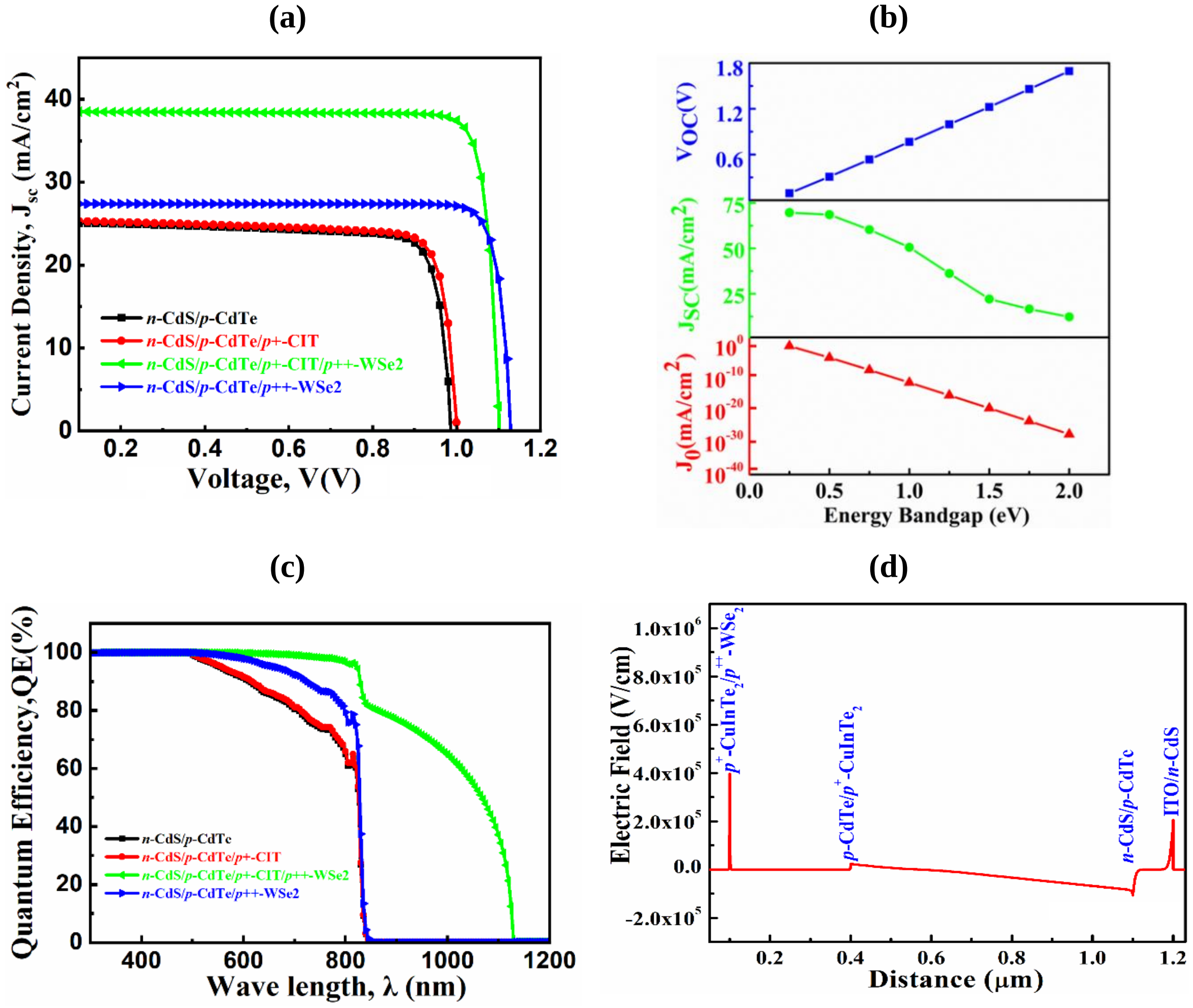

**Figure 2:** (a) J-V characteristic curves of *n*-CdS/*p*-CdTe/$p^+$-CIT/$p^{++}$-$WSe_2$ solar cell, (b) Bandgap dependent $V_{OC}$, $J_{SC}$ and $J_0$ of a solar cell at AM 1.5G, (c) QE spectra of *n*-CdS/*p*-CdTe/$p^+$-CIT/$p^{++}$-$WSe_2$ cell and (d) Built-in electric field generated at the interfaces of the *n*-CdS/*p*-CdTe/$p^+$-CIT/$p^{++}$-$WSe_2$ heterojunction PV cell.

Now, with the inclusion of second $p^+$-CIT absorber, the designed *n*-CdS/*p*-CdTe/$p^+$-CIT/$p^{++}$-$WSe_2$ cell provides the open circuit voltage of 1.102 Volt which is ~0.02 volt less than that of n-CdS/*p*-CdTe/$p^+$-$WSe_2$ heterostructure. However, it produces the highest current density of 38.5 mA/cm$^2$ that is 25% more current density than the previous configuration. The device exhibits an efficiency of 37.46% with a FF of 88.30%. The small reduction in $V_{OC}$ is reasonable as a relatively narrow bandgap CIT (1.1 eV) layer has been added as an active absorber in the device that might increases the dark current as shown in Figure 2(b). On the contrary, the significant rise in $J_{SC}$ can be interpreted in terms of the reduction of the minority carrier recombination velocity by the BSF. The quantum efficiency (QE) can explain further the origin of huge current in the device.

The quantum efficiency (QE) of the corresponding solar cells are shown in Figure 2(c). The *n*-CdS/*p*-CdTe solar PV cell exhibits almost 100% QE from 300 nm to 500 nm wavelength. After 500 nm the QE falls very sharply and becomes 0 at ~845 nm wavelength. Therefore, this cell can absorb light from 500 to ~845 nm and cannot absorb completely light with wavelength higher than 845 nm. The *n*-CdS/*p*-CdTe/$p^+$-$WSe_2$ which is essentially the first solar cell with added BSF layer to help reduce the recombination of the carrier pairs thus creating higher QE around the 500 to 845 nm wavelength range. Notice that the wavelength that the absorber can absorb is still limited to 845 nm. However, *n*-CdS/*p*-CdTe/$p^+$-CIT/$p^{++}$-$WSe_2$ solar cell can absorb light from 300 nm to 1125 nm as indicated in the figure. This is because of the combined absorption of the CdTe and CIT layers. The bandgap of CdTe is 1.5 eV and therefore its cut off wavelength is around 845 nm whereas CIT has a bandgap of 1.1 eV with a cut-off wavelength of around 1125 nm. The addition of $p^+$-CIT in the structure moves BSF to the $P^+$-CIT/$p^{++}$-$WSe_2$ interface. Therefore, the powerful built-in electric field generated in the interface helps separating the generated the electron-hole pairs by the absorption of photons with low energy ($1.5 \geq h\nu \geq 1.1$ eV) in the CIT layer. The added BSF layer helps in the reduction of recombination of the carrier pairs hence QE increases significantly for this additional 845 nm to 1125 nm wavelength. In this solar cell, QE is approximately 100% to 95% for 300 nm to 845 nm. The quantum efficiency drops gradually for the wavelengths higher than 845 nm. It indicates that this combination of solar cells can absorbs lower energy photons thus enhancing the PCE of the CdTe solar cell.

However, CIT layer alone cannot increase current in *n*-CdS/*p*-CdTe/$p^+$-CIT device structure. This structure gives a $V_{OC}$ of about 1 V and $J_{SC}$ of 25.4 mA/cm$^2$ with $\Delta V_{OC}$ and $\Delta J_{SC}$ of 0.02 V and 0.2 mA/cm$^2$. Therefore, *n*-CdS/*p*-CdTe/$p^+$-CIT structure does not show any promising change in voltage or current density. The *n*-CdS/*p*-CdTe/$p^+$-CIT solar cell with the added absorber layer of $p^+$-$CuInTe_2$ shows almost the exact characteristics as *n*-CdS/*p*-CdTe solar cell in regards to the wavelength of light and the quantum efficiency. This is because *p*-CdTe/$p^+$-CIT heterojunction builds almost no electric field at the junction as shown in Figure 2(d) which has been drawn from C-V analysis. Therefore, this interface does not act as a BSF and cannot provides any $V_{OC}$ as well as it cannot separate any electron-hole pairs produced when low energy photon get absorbed in the CIT layer.

The following sections recapitulate the effects of the alteration on the physical parameters of various constituent layers of the *n*-CdS/*p*-CdTe/$p^+$-CIT/$p^{++}$-$WSe_2$ thin film solar cells.

## 3.2 Role of CdTe absorber in performance

Figure 3(a) shows the effects of variation in the width of the *p*-CdTe absorber on solar cell parameters. The breadth of the CdTe absorber has been changed from 200 to 1200 nm. With the variation of the thickness of the absorption layer, the current density and efficiency are slightly changed. Therefore, the thickness of CdTe has no major impact on the CdTe solar cell. When the thickness is raised from 200 nm, the current density starts to increase up to 600 nm and after that, the current density and $V_{OC}$ falls slightly. The depth of the layer is kept at 700 nm for the entire study from the experimental point of view.

Figure 3(b) depicts the function of doping density on the output parameters such as $V_{OC}$, $J_{SC}$, FF, and PCE of the CdTe solar cell. Acceptor doping in CdTe has been varied from $10^{11}$ to $10^{20}$ cm$^{-3}$. It is perceived in the figure that there are very few changes in $V_{OC}$, $J_{SC}$, FF, and PCE up to the doping value of $10^{16}$ cm$^{-3}$. There are also some variations in FF but not that significant changes are monitored. At any higher value than $10^{16}$ cm$^{-3}$, the current density and efficiency drop sharply. At a doping concentration level of $10^{19}$ cm$^{-3}$, current density becomes 16.24 mA/cm$^2$ and the efficiency becomes 15.77%. The decrease in $J_{SC}$ and efficiency at higher doping may be attributed to the losses by radiative recombination in the solar PV cell [25]. Therefore, the doping concentration for this layer has been selected at $10^{16}$ cm$^{-3}$ considering present manufacturing standards.

(a) (b) (c)

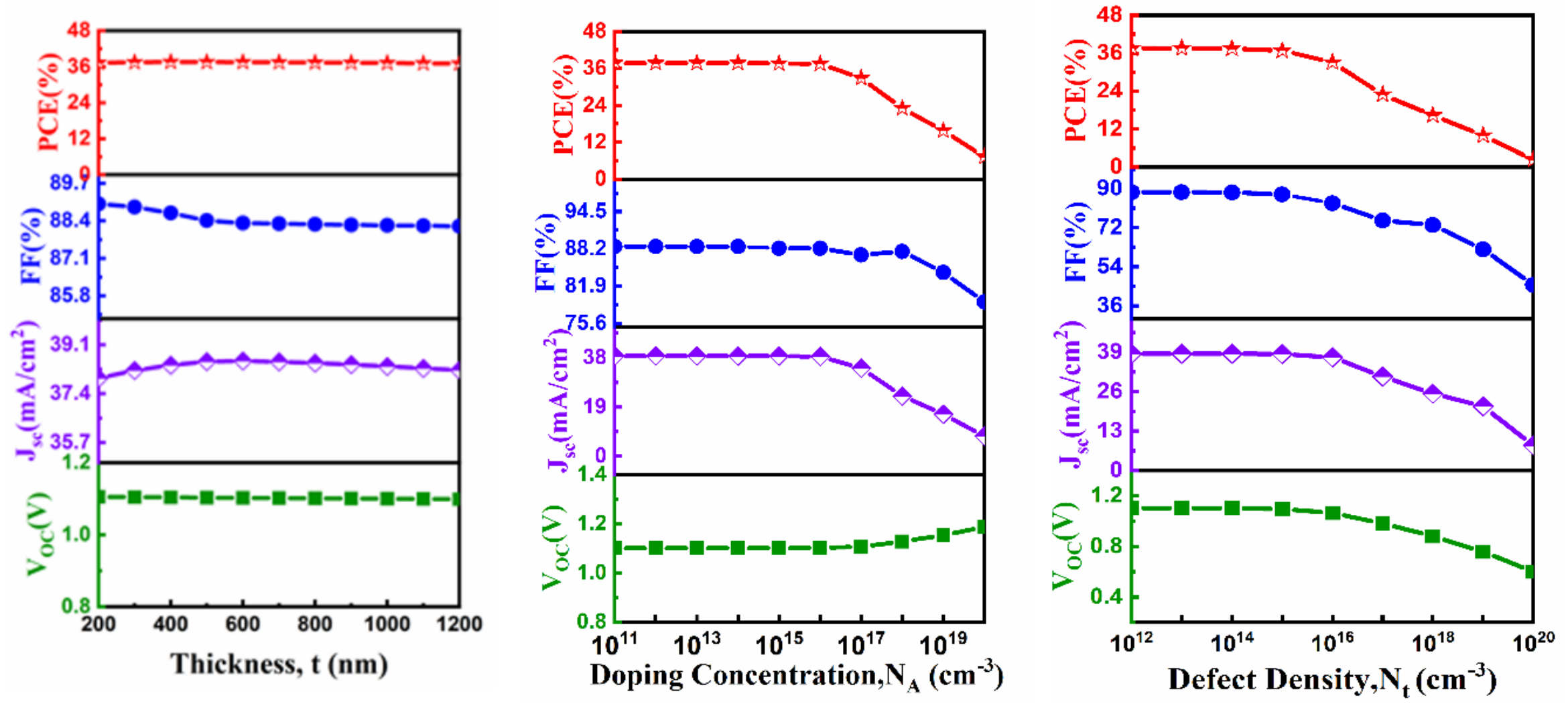

**Figure 3:** Variation of PV performance with CdTe absorber layer (a) Depth, (b) Doping, and (c) Defect density of *n*-CdS/*p*-CdTe/*p*$^+$-CIT/*p*$^{++}$-WSe$_2$ solar cell.

Figure 3(c) reveals the influence of defect density on different parameters such as $V_{OC}$, $J_{SC}$, FF, and PCE of the CdTe device. The defect density in the semiconductor has been switched from $10^{12}$ to $10^{20}$ cm$^{-3}$. It is visualized from the figure that the efficiency, current density, and voltage can vary significantly. Defect density up to $10^{15}$ cm$^{-3}$ gives a stable value in $V_{OC}$, $J_{SC}$, FF, and PCE. With the rise in defect density above $10^{15}$ cm$^{-3}$, $V_{OC}$, $J_{SC}$, FF, and PCE decrease notably. At a defect density of $10^{17}$ cm$^{-3}$, $V_{OC}$ drops from 1.102 volt to 0.98 volt also current density decrease about 7.6 mA/cm$^2$ from the values of defect density of $10^{14}$ cm$^{-3}$. A similar decrement in FF and PCE are also observed for increasing defect density. These falls in solar cell parameters are due to the decrease in carrier lifetime and hence rise in recombination current with defects [26]. So, the bulk defects of the CdTe semiconductor seriously affects the efficiency of the solar cell. For further simulations, a bulk defect of the order of $10^{14}$ $cm^{-3}$ has been taken as an optimum value.

### 3.3 $CuInTe_2$ layer impact on performance

Figure 4(a) shows the upshot of the depth of the $CuInTe_2$ absorber layer in CdTe solar cell. The width of the CIT layer has been changed from 0 to 400 nm. It is noted from the graph that $V_{OC}$ of the device marginally falls with the increasing depth of the CIT absorber. But the current density, $J_{SC}$ rises with the growth in the $CuInTe_2$ layer thickness. FF decreases slightly with the incremental values of thickness. And therefore, the PCE of the CdTe solar cell climbs up with the increasing width of $CuInTe_2$. This is a CdTe-based thin film solar cell, and the purpose of the use of CIT semiconductor is to absorb higher wavelength light thus enhancing current and thus PCE of the cell. Therefore, a thin CIT semiconducting layer has been considered in this simulation.

Figure 4(b) shows the influence of the acceptor doping in the CIT semiconductor in output parameters of the CdTe solar cell. With the increment of doping density all the four parameters of our discussion increase gradually. The $V_{OC}$ of the CdTe solar cell mounts with doping in the CIT layer. The value of $V_{OC}$ is 0.978 V at a doping of $10^{17}$ $cm^{-3}$, which increases to 1.102 V at a doping of $5\times10^{19}$ $cm^{-3}$. The open circuit voltage, $V_{OC}$ relies on the acceptor concentration, $N_A$ by the following formula [27]

$$V_{OC} = \frac{K_B T}{q} ln\left[\frac{(N_A+\Delta n)\Delta n}{n_i^2}\right] \quad (8)$$

where, $\frac{K_B T}{q}$ represents the thermal voltage, Δn denotes the excess carriers and $n_i$ stands for the intrinsic carrier density. Therefore, the climb in $V_{OC}$ of the solar cell is reasonable with the increment in doping in the CIT layer.

However, $J_{SC}$ increases with doping in CIT layer in an unusual fashion which might be attributed to the rise in built-in electric field in CIT/$WSe_2$ interface that enhances the spliting of photgenerated carriers in the CIT layer [22, 24, 28]. The fill factor, FF of the solar cel also increase by doping in CIT layer. FF is related to $V_{OC}$ and diode ideality factor, n by $FF = \frac{v_{OC} - \ln(v_{OC} - 0.72)}{v_{OC}}$, where $v_{OC} = \frac{V_{OC}}{nK_BT/q}$ [11]. As the $V_{OC}$ of the devices rises with doping the FF of the solar device also enhances [29]. As the three of the parameters rise with doping in CIT layer, the PCE of the solar cell augments. But in our modeled CdTe solar cell, we have considered the doping value of $5\times10^{19}$ $cm^{-3}$. Because of the lower values of all four parameters at low doping concentrations, those lower values have not been used. Although the whole performance of the solar cell increases significantly with the increment of doping density, at doping levels higher than $5\times10^{19}$ $cm^{-3}$, the CdTe solar cell may suffers from low carrier lifetime owing to the rise in radiative recombination and also higher doping may reduce the carrier mobility which in turn may reduce carrier diffusion length and thus decrease short circuit current in the device [25-26]. Therefore, the thickness and doping in CIT layer should be optimized for highest outcome of the *n*-CdS/*p*-CdTe/$p^+$-CIT/$p^{++}$-$WSe_2$ photovoltaic cell.

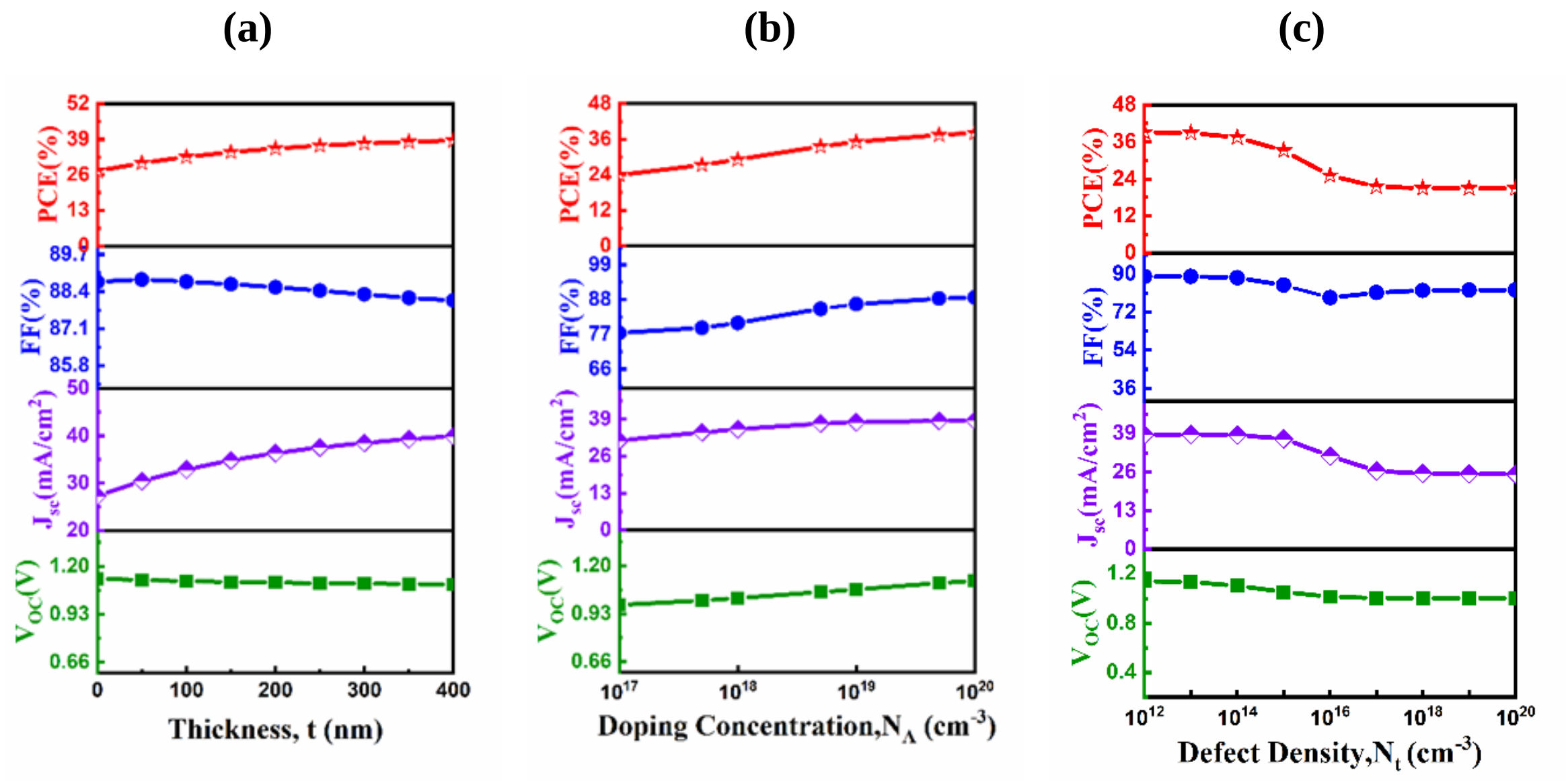

**Figure 4:** Variation of PV performance with $CuInTe_2$ absorber layer (a) Depth, (b) Doping, and (c) bulk defects of *n*-CdS/*p*-CdTe/$p^+$-CIT/$p^{++}$-$WSe_2$ solar cell.

Figure 4(c) expresses the role of defect density of the $CuInTe_2$ semiconductor layer in solar cell parameters. The bulk defect density has been switched in the range of $10^{12}$ to $10^{20}$ $cm^{-3}$ to understand the possible effects on various PV outcomes. The increasing value of defects lowers all four parameters discussed in this simulation. Also considering higher values of defects for possible outcomes, we observe that efficiency drops from 37.46% to 33.18% when defect density order increased from $10^{14}$ to $10^{15}$ $cm^{-3}$ and the decrement rate in efficiency, current density, and voltage is more with higher order of defects which indicates that defect density of CIT layer significantly affects the efficiency and overall performance of the CdTe solar cell. The proliferation in dark current due to Shockley–Read–Hall (SRH) recombination which shortens the carrier lifetime is the main reason of decreasing the solar cell performance with increasing defects in CIT layer [30]. A bulk defects of the order of $10^{14}$ $cm^{-3}$ has been considered as a safe value for the present simulation.

### 3.4 Role of CdS window and $WSe_2$ BSF Layers

#### *3.4.1 CdS layer impact*

Figure 5(a) presents the variation in solar cell parameters, namely $V_{OC}$, $J_{SC}$, FF, and PCE, of the proposed photovoltaic (PV) cell by the variation of the CdS window's thickness. The depth of the window layer is altered from 50 to 300 nm. With the change in thickness, there is a very slight noticeable shift in the solar cell's parameters. The PCE of the solar cell is nearly constant owing to the changes in the width of the window layer. In CdS layer with a realistic defect of $10^{14}$ $cm^{-3}$, the carrier lifetime and hence the length of diffusion of electrons and holes are high enough to be collected efficiently before their recombination in the investigated range of thickness [31]. Therefore, all the parameters are almost constant. However, the simulation has been carried out with an optimum width of the window layer of 100 nm.

Figure 5(b) illustrates the effects of doping density of the CdS layer on the output of CdTe solar cell. The donor doping in CdS semiconductor has been altered in the span of $10^{15}$ to $10^{21}$ $cm^{-3}$. It is seen from the figure that there is almost no change in output parameters due to the rise of doping concentration. However, at a very elevated doping the device performance may degrade due to the advancement in dark current by the increase in radiative recombination [32]. Therefore, the doping in CdS semiconductor is kept at $10^{18}$ $cm^{-3}$ for the simulation.

Figure 5(c) shows the dominance of defects in the CdS window layer on the output parameters in the CdTe cell. The bulk defect density in CdS semiconductor has been changed in the order of $10^{12}$ to $10^{19}$ $cm^{-3}$. It can be observed in the figure that the current density and efficiency of the solar cell stays quite the same up to defects of $10^{16}$ $cm^{-3}$. After that level, we observed a massive drop in $J_{SC}$, FF, and PCE. When defect density increases from $10^{16}$ to $10^{18}$ $cm^{-3}$, the efficiency falls from 37.38% to 36.7%. And after that $J_{SC}$ and hence PCE of the device significantly drop. The carrier lifetime and hence the diffusion length of photocarriers are reduced due to defect related SRH recombination. And as a consequence, the short circuit current and therefore the PCE significantly decreases [31-32]. So, the defect density has been kept at $10^{14}$ $cm^{-3}$ which is a feasible value in current solar cells manufacturing technology.

(a) (b) (c)

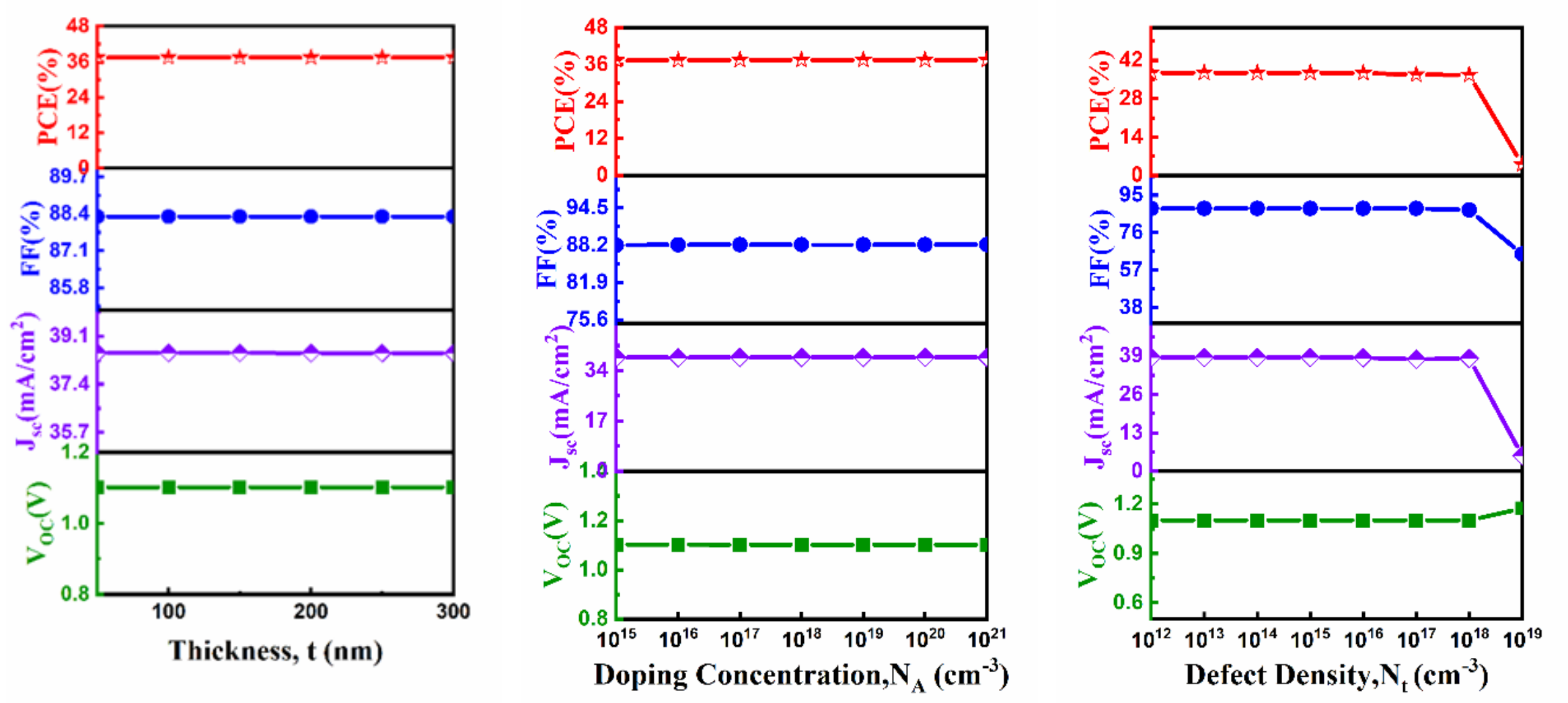

**Figure 5:** Variation of PV performance with n-CdS window layer (a) Depth, (b) Doping, and (c) Defect density of *n*-CdS/*p*-CdTe/$p^+$-CIT/$p^{++}$-$WSe_2$ solar cell.

### *3.4.2 $WSe_2$ BSF impact*

In this section, the impacts of different parameters of $p^{++}$-$WSe_2$ back surface on the output of CdTe thin film PV cell have been enumerated in details.

Figure 6(a) shows the role of the breadth of the $WSe_2$ BSF semiconductor in CdTe solar cell performances. The thickness of $WSe_2$ has been altered from 50 to 400 nm. With the change in the width of the BSF layer, there are very little changes in the four output parameters which indicates there is very little effect of thickness on the PV outcomes of the CdTe PV device. The main purpose of the use of this layer is to help in carrier separation in absorber layer thus increasing the PCE of the solar cell. This layer doesn't contribute directly to the absorption but helps the absorption of light of higher wavelengths. Therefore, the $WSe_2$ BSF layer thickness has been kept at 100 nm.

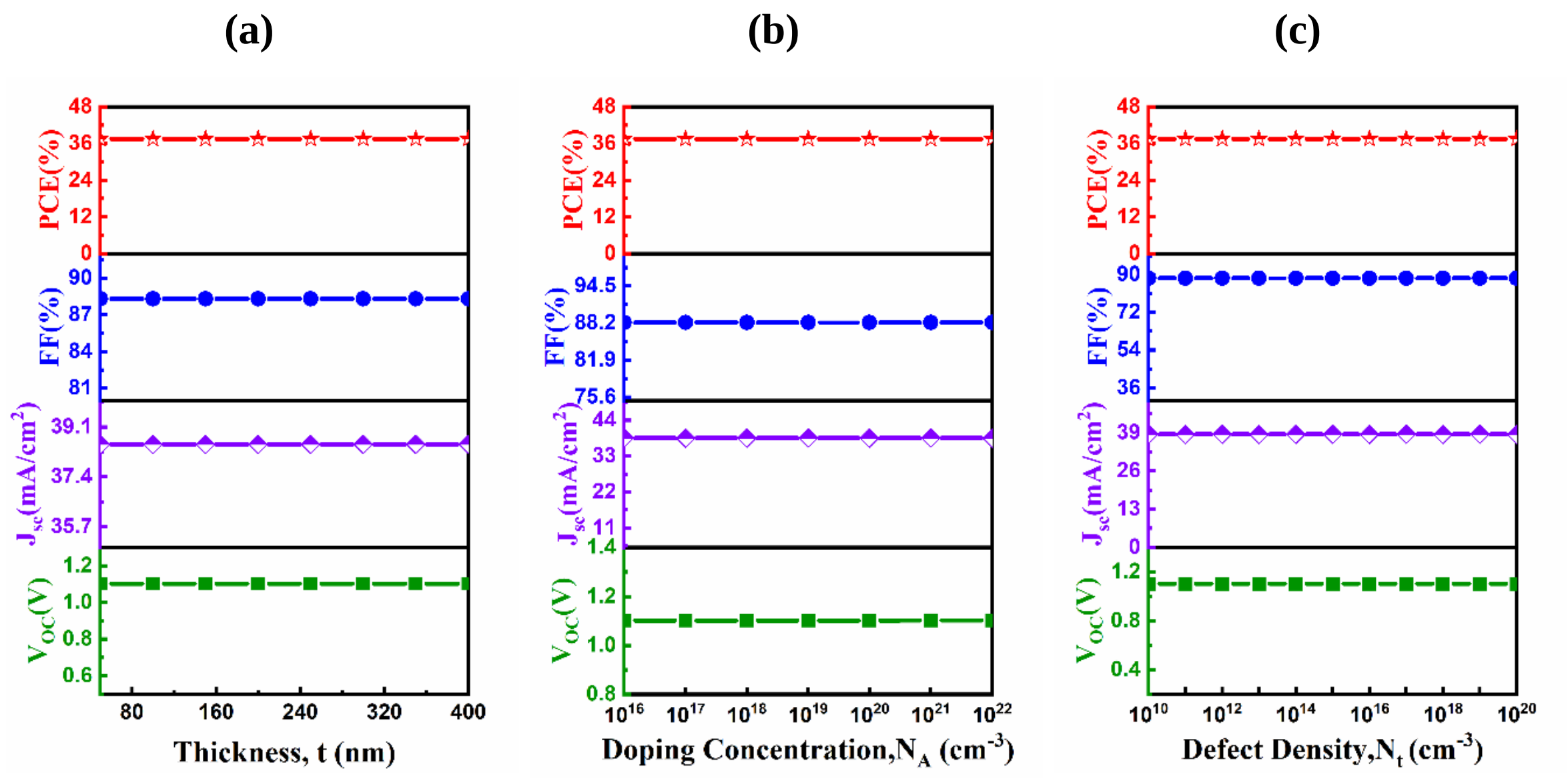

**Figure 6:** PV performance with $WSe_2$ back surface field layer (a) Depth, (b) Acceptor density, and (c) Bulk defects variation of *n*-CdS/*p*-CdTe/$p^+$-CIT/$p^{++}$-$WSe_2$ heterojunction solar cell.

Figure 6(b) reveals the dominance of the doping density of $WSe_2$ semiconducting layer on $V_{OC}$, $J_{SC}$, FF, and PCE of the CdTe PV device. The doping density in the BSF has been varied from $10^{16}$ to $10^{20}$ $cm^{-3}$ and the effect is very negligible. As there is no noteworthy change in performance, we kept this BSF layer concentration to $10^{20}$ $cm^{-3}$.

Figure 6(c) illustrates the role of the defect density of $WSe_2$ BSF Layer in CdTe solar cells. The volume defect density of the BSF has been varied from $10^{10}$ to $10^{20}$ $cm^{-3}$ and there is no change in the efficiency and other parameters of the solar cell. So, the optimum defect density level has been kept to $10^{15}$ $cm^{-3}$.

### 3.5 Temperature effect on CdTe solar cell

The working temperature of the solar cells has massive influence on the operation of the devices. The density of states (DOS) in the conduction and valance bands of a semiconductor depend on temperature [20]. The thermal velocity of the carriers are also temperature dependent. The semiconductor bandgap also reduces with the increase in temperature [33]. Therefore, the cell temperature changes the device performances.

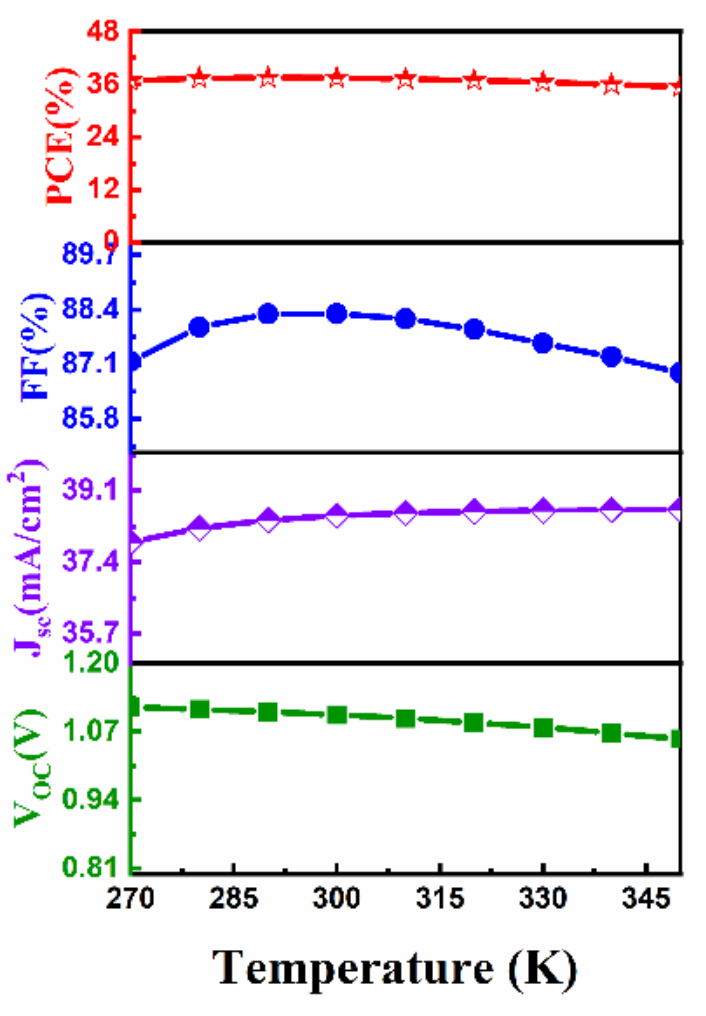

**Figure 7:** Effect of operating temperature on the output of *n*-CdS/*p*-CdTe/$p^{+}$-CIT/$p^{++}$-$WSe_2$ PV cell.

Figure 7 depicts the role of temperature on the performance of CdTe photovoltaic device in the span of 270 to 350 K. With the increment of temperature, the voltage, $V_{OC}$ gradually falls. This is happened owing to the reduction in semiconductor bandgap that consequents in the increase in dark current as shown in Figure 2(b). On the contrary, the short circuit current, $J_{SC}$ enhances gradually as the bandgap diminishes with temperature. But FF increases by 1% or 2% at room temperature or around 300 to 310 K. The increment in current and decrement in voltage balances the PCE of the solar cell with temperature. So, PCE doesn't change notably with the investigated temperature range of the CdTe solar cell indicating the stability of the device.

### 3.6 Resistance effect on CdTe solar cell

The various resistances such as series and shunt resistances in solar cell diode has significant impact on solar cell performances. These resistances are originated from semiconductor bulk body, different types of contacts and leakage currents due to fabrication defects in solar cells [29]. Figure 8 depicts the impacts of series and shunt resistances on the output performances of CdTe solar cells.

Figure 8(a) shows the series resistance's effect on the voltage, current, field, and efficiency of $n$-CdS/$p$-CdTe/$p^+$-CIT/$p^{++}$-$WSe_2$ heterojunction PV cell considering the shunt resistance infinity. It is observed from the figure that both the device voltage and current are almost constant with the variation of series resistance. It is the FF which is severely affected by the series resistance and therefore efficiency of the CdTe solar cell falls. When series resistance is 0 $\Omega.cm^2$, the PCE of the CdTe device is 37.46% with an FF of 88.3%. With the increment of the value of series resistance the FF decreases gradually from 88.3% to 72.20% at series resistance of 5 $\Omega.cm^2$ and hence PCE falls to 30.62%. Therefore, the series resistance should be optimized in practical $n$-CdS/$p$-CdTe/$p^+$-CIT/$p^{++}$-$WSe_2$ PV cell to gain the highest PCE of the solar cell.

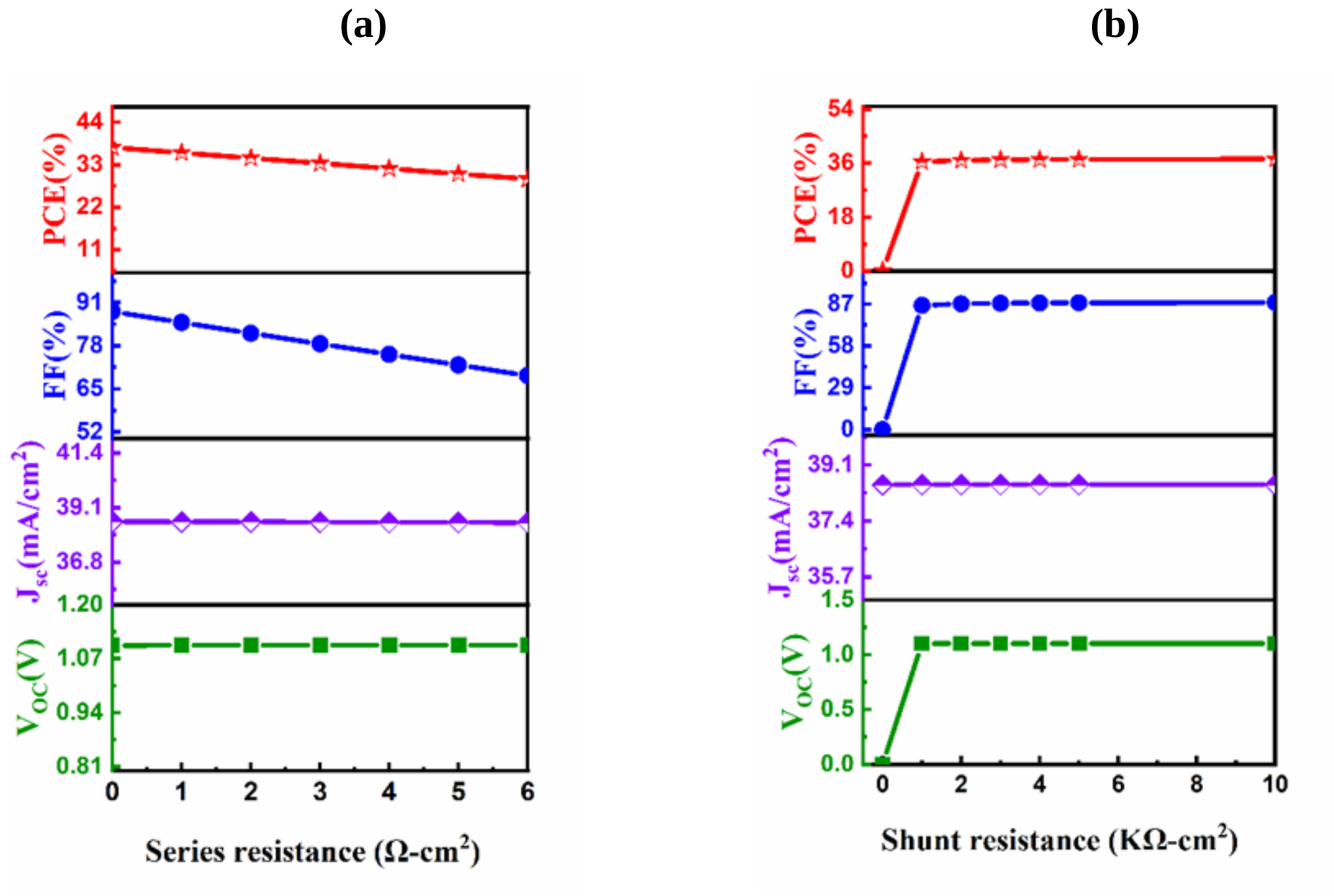

**Figure 8:** Effect of (a) Series and (b) Shunt resistances of *n*-CdS/*p*-CdTe/$p^+$-CIT/$p^{++}$-$WSe_2$ PV cell.

Figure 8(b) shows the shunt resistance's effect on the voltage, current, field, and efficiency of *n*-CdS/*p*-CdTe/$p^+$-CIT/$p^{++}$-$WSe_2$ solar cell taking series resistance as an ideal value i.e. 0 Ω. The shunt resistance in the figure has been varied from 0 to 10 kΩ.cm$^2$. The lower value of shunt resistance significantly affects the $V_{OC}$ and FF and hence the efficiency of the device. At shunt resistance 0 Ω.cm$^2$, there is only short circuit current density does exist, therefore, the efficiency of the *n*-CdS/*p*-CdTe/$p^+$-CIT/$p^{++}$-$WSe_2$ device becomes zero. When the shunt resistance increases from zero to the higher value, the $V_{OC}$ and FF increases and reach at maximum values at a shunt resistance of infinity. It is noticed from the figure that $V_{OC}$ and FF are 1.101 V and 85.97% at a shunt resistance of 1 k Ω.cm$^2$ that provide an efficiency of 36.45%. The $V_{OC}$ and FF are 1.102 V and 87.84% at a shunt resistance of 5 k Ω.cm$^2$ that yield a PCE of 37.26%. Therefore, it can be summarized that the value of shunt resistance should as high as possible in order to gain a maximum efficiency of the *n*-CdS/*p*-CdTe/$p^+$-CIT/$p^{++}$-$WSe_2$ thin film heterojunction PV cell.

### 3.7 Photon recycling impact on CdTe solar cell

The bandgap of CIT is 1.1 eV and theoretical short circuit current limit for this absorber is ~46 $mA/cm^2$ as depicted in Figure 2(b). Therefore, the current in CdTe-CIT solar cell could be further improved if a thicker CIT layer is used. However, the doping in the layer is slightly high, therefore, the increase in CIT layer thickness may have a negative effect in the solar cell as the dark current will increase due to increase in radiative recombination with higher doping [32]. However, the current in the solar cell might be increased with the use of a Bragg Reflector (BR) with which nearly 100% reflectance can be gained for a limited range of wavelength by exploiting layer by layer stacking of materials having dissimilar refractive indices [34].

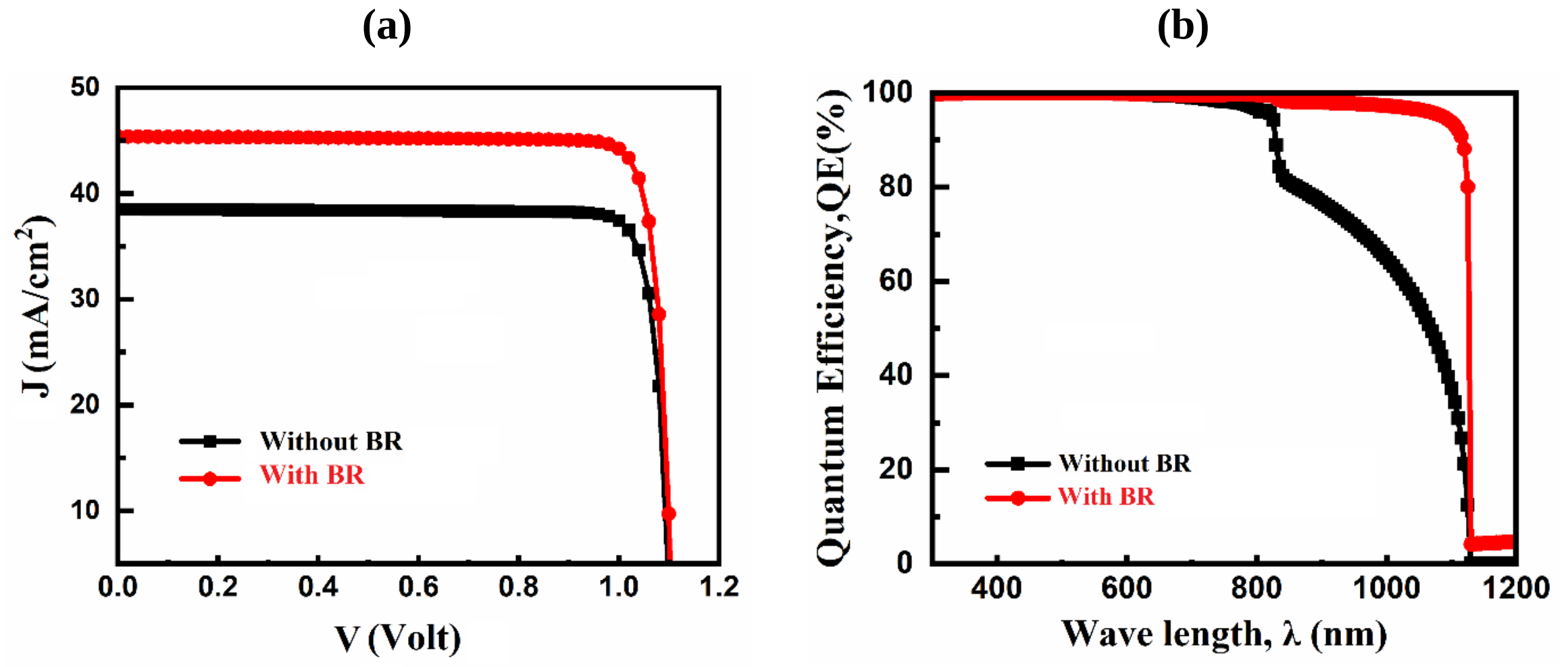

**Figure 9:** (a) Current (J) –voltage (V) and (b) quantum efficiency (QE) spectra of *n*-CdS/*p*-CdTe/$p^+$-CIT/$p^{++}$-$WSe_2$ solar cell with and without Bragg reflection.

Figure 9(a) displays the J-V curve of *n*-CdS/*p*-CdTe/$p^+$-CIT/$p^{++}$-$WSe_2$ thin film PV cell with and without BR. Bragg reflector having 98% back and 95% front reflection has been considered for the calculation. With the use of this BR, there is a significant improvement in current density and quantum efficiency. Here, the current density has raised from 38.5

mA/cm$^2$ to 45.4 mA/cm$^2$ which is 18% higher which in turn significantly increases the PCE of the device from 37.46 to 44.32%.

Figure 9(b) depicts the wavelength vs. quantum efficiency (QE) spectra of the *n*-CdS/*p*-CdTe/$p^+$-CIT/$p^{++}$-$WSe_2$ thin film PV device with and without BR. The figure indicates that the quantum efficiency of the *n*-CdS/*p*-CdTe/$p^+$-CIT/$p^{++}$-$WSe_2$ PV cell highly escalated in the longer wavelength region due to the use of Bragg reflector. Without BR, quantum efficiency falls notably for higher wavelengths than 845 nm but with BR, the solar cell retains quantum efficiency from 80 to 98% up to wavelength 1125 nm. The higher QE ensures the higher $J_{SC}$ of the devices and hence the PCE of the device significantly increases.

## 4. Conclusions

In this work, we establish a highly promising *n*-CdS/*p*-CdTe/$p^+$-CIT/$p^{++}$-$WSe_2$ solar cell. In the structure, CdTe and $CuInTe_2$ layers have thicknesses of 700 and 300 nm, respectively. CdS window and $WSe_2$ BSF layers both have the thickness of 100 nm. The addition of $CuInTe_2$ with a bandgap of 1.1 eV enables the solar cell to absorb light up to the wavelength of 1125 nm. To improve the overall efficiency, $p^{++}$-$WSe_2$ BSF layer has been added which increase the efficiency of the CdTe PV cell from 20.43 to 37.46%. To further improve the efficiency, Bragg reflector has been considered and the efficiency CdTe solar cell enhances up to 44.31%. The simulation is performed taking the optimum doping and bulk defects and with other standard parameters. Therefore, the manufacture of this solar cell is feasible in future.

## Acknowledgements

The authors indebted to Prof. Marc Burgelman, University of Gent, Belgium, for supplying SCAPS simulation software.

**Corresponding Author:**

*E-mail: jak_apee@ru.ac.bd (Jaker Hossain).

**Disclosures:** The authors declare no competing financial interest.

**Data availability:** Simulation details and associated data are available free of charge from authors upon reasonable request.

## References

[1] M. G. Reyes-Banda, E. Regalado-Perez, M. I. Pintor-Monroy , C. A. Hernandez-Gutierrez, M. A. Quevedo-Lopez, X. Mathew, Effect of Se diffusion and the role of a thin CdS buffer layer in the performance of a CdSe/CdTe solar cell, Superlattices Microstruct. 133 (2019) 106219.

[2] X. Li, D. Xiao, L. Wu, D. Wang, G. Wang, D. Wang, CdTe thin film solar cells with copper iodide as a back contact buffer layer, Solar Energy 185 (2019) 324–332

[3] M. A. Green, E. D. Dunlop, J. Hohl-Ebinger, M. Yoshita, N. Kopidakis, X. Hao, Solar cell efficiency tables (version 56), Prog. Photovolt. Res. Appl. 28 (2020) 629–638.

[4] P. Baruch, A. D Vos, P. T. Landsberg, and J. E. Parrott, On some thermodynamic aspects of photovoltaic solar energy conversion, Sol. Energy Mater. Solar Cells 36 (1995) 201-222.

[5] M. I. Hossain, W. Qarony, S. Ma, L. Zeng, D. Knipp, Y. H. Tsang, Perovskite/Silicon Tandem Solar Cells: From Detailed Balance Limit Calculations to Photon Management, Nano-Micro Lett. 11 (2019) 58.

[6] N. Romeo, Solar cells made by chalcopyrite materials, Jpn. J. Appl. Phys. **19** (1980) 5-13.

[7] J. J. Loferski, Theoretical considerations governing the choice of the optimum semiconductor for photovoltaic solar energy conversion, J. Appl. Phys. 27 (1956) 777-784

[8] I. M. Dharmadasa and A. E. Alam, How to achieve efficiencies beyond 22.1% for CdTe-based thin-film solar cells, Energies 15 (2022) 9510.

[9] C. Li, F. Wang, Y. Chen, L. Wu, J. Zhang, W. Li, X. He, B. Li, and L. Feng, Characterization of sputtered CdSe thin films as the window layer for CdTe solar cells, Materi. Sci. Semicond. Process. 83 (2018) 89–95

[10] I. M. Dharmadasa, A. E. Alam, A. A. Ojo and O. K. Echendu, Scientific complications and controversies noted in the field of CdS/ CdTe thin film solar cells and the way forward for further development, J. Mater. Sci.: Mater. Electron. 30 (2019) 20330–20344.

[11] M. A. H. Pappu, A. Kuddus, B. K. Mondol, A.T. Abir, J. Hossain, Design of n-CdS/p-$CuInTe_2$/p+-$MoS_2$ thin film solar cell with a power conversion efficiency of 34.32%, Optics Continuum 2 (2023) 942-955.

[12] T. Mise, and T. Nakada, Low temperature growth and properties of Cu–In–Te based thin films for narrow bandgap solar cells, Thin Solid Films 518 (2010) 5604-5609.

[13] G. Jia, B. Liu, K. Wang, C. Wang, P. Yang, J. Liu, W. Zhang, R. Li, S. Zhang, and J. Du, $CuInTe_2$ Nanocrystals: shape and size control, formation mechanism and application, and use as photovoltaics, Nanomaterials 9 (2019) 409.

[14] N. A. Patil, M. Lakhe, and N. B. Chaure, Characterization of $CuInTe_2$ thin films deposited by electrochemical technique, AIP Conference Proceedings 1447 (2012) 1073–1074.

[15] J. Hossain, B. K. Mondal and S. K. Mostaque, Computational investigation on the photovoltaic performance of an efficient GeSe-based dual-heterojunction thin film solar cell, Semicond. Sci. Technol. 37 (2022) 015008.

[16] Z. Zhang, P. Chen, X. Yang, Y. Liu, H. Ma, J. Li, B. Zhao, J. Luo, X. Duan and X. Duan, Ultrafast growth of large single crystals of monolayer $WS_2$ and $WSe_2$, Natl Sci. Rev. 7 (2020) 737–44.

[17] A. Kuddus, A. B. M. Ismail and J. Hossain, Design of a highly efficient CdTe-based dual-heterojunction solar cell with 44% predicted efficiency, Solar Energy 221 (2021) 488–501.

[18] M. Burgelman, J. Verschraegen, S. Degrave, P. Nollet, Modeling thin-film PV devices, Prog. Photovoltaics Res. Appl. 12 (2004) 143–153.

[19] B. K. Mondal, S. K. Mostaque, and J. Hossain, Unraveling the effects of a GeSe BSF layer on the performance of a CuInSe2 thin film solar cell: a computational analysis, Optics Continuum 2 (2023) 428-440.

[20] A. T. Abir, B. K. Mondal and J. Hossain, Exploring the potential of GeTe for the application in Thermophotovoltaic (TPV) cell. arXiv (Preprint), 10.48550/arXiv.2301.08896

[21] P. Baruch, A. De Vos, P. T. Landsberg, and J. E. Parrott, On some thermodynamic aspects of photovoltaic solar energy conversion, Sol. Energy Mater. Solar Cells, 36 (1995) 201-222.

[22] J. G. Fossum, Physical operation of back-surface-field silicon solar cells, IEEE Trans. Electr. Devices 24 (1977) 322-325.

[23] S. R. Dhariwal and Arun P. Kulshreshth, Theory of back surface field silicon solar cells, Solid-State Electronics 24 (1981) 1161-l 165.

[24] A. Sinha and S. K. Chaitopadhyaya, Effect of back surface field on photocurrent in a semiconductor junction, Solid-State Electronics 21 (1978) 943-951.

[25] A. Kanevce and T. A. Gessert, Optimizing CdTe solar cell performance: impact of variations in minority-carrier lifetime and carrier density profile, IEEE J. Photovoltaics 1 (2011) 99-103

[26] M. M. A. Moon, M. F. Rahman, M. Kamruzzaman, J. Hossain, A. B. M. Ismail, Unveiling the prospect of a novel chemical route for synthesizing solution-processed CdS/CdTe thin-film solar cells, Energy Reports 7 (2021) 1742-1756.

[27] R. A. Sinton and A. Cuevas, Contactless determination of current–voltage characteristics and minority-carrier lifetimes in semiconductors from quasi-steady-state photoconductance data, Appl. Phys. Lett. 69 (1996) 2510-2512.

[28] J. Hossain, M. M. A. Moon, B. K. Mondal, M. A. Halim, Design guidelines for a highly efficient high-purity germanium (HPGe)-based double-heterojunction solar cell, Optics and Laser Technology 143 (2021) 107306.

[29] M. C. Islam, B. K. Mondal, T. Ahmed, M. A. H. Pappu, S. K. Mostaque and J. Hossain, Design of a highly efficient n-CdS/p-$AgGaTe_2$/$p^+$-SnS double-heterojunction thin film solar cell, Eng. Res. Express 5 (2023) 025056.

[30] D. Kuciauskas, J. Moseley, and C. Lee, Identification of recombination losses in CdSe/CdTe solar cells from spectroscopic and microscopic time-resolved photoluminescence, Solar RRL 5 (2021) 2000775.

[31] M. S. Hossen, A. T. Abir, and J. Hossain, Design of an Efficient $AgInSe_2$ Chalcopyrite-Based Heterojunction Thin-Film Solar Cell, Energy Technol. 11 (2023) 2300223.

[32] B. Das, I. Aguilera, U. Rau, and T. Kirchartz, Effect of doping, photodoping, and bandgap variation on the performance of perovskite solar cells, Adv. Optical Mater. 10 (2022) 2101947.

[33] A. Kuddus, M. F. Rahman, J. Hossain, and A. B. M. Ismail, Enhancement of the performance of CdS/CdTe heterojunction solar cell using $TiO_2$/ZnO bi-layer ARC and $V_2O_5$ BSF layers: A simulation approach, Eur. Phys. J. Appl. Phys. 92, 20901 (2020).

[34] M. Z. Shvarts, O. I. Chosta, I. V. Kochnev, V. M. Lantratov, V. M. Andreev, Radiation resistant AlGaAs/GaAs concentrator solar cells with internal Bragg reflector, Sol. Energy Mater. Sol. Cells 68 (2001) 105-122.